\documentclass[oneside,a4paper,12pt,nofootinbib,floatfix]{article}

\usepackage{graphicx, wasysym}
\usepackage{amssymb} 

\begin{document}

\thispagestyle{empty}
 {\flushleft \LARGE \bf Constraints on Cosmological Dark Matter \\[0.2mm] 
Annihilation from the Fermi-LAT Isotropic \\[2.5mm] 
Diffuse Gamma-Ray Measurement}
%\title{Constraints on Cosmological Dark Matter Annihilation from the Fermi-LAT Isotropic Diffuse Gamma-Ray Measurement}

\noindent\footnotetext[0]{\bf Corresponding authors:\newline 
Jan Conrad: {\rm conrad@fysik.su.se} \newline 
Michael Gustafsson: {\rm michael.gustafsson@pd.infn.it}  \newline 
Alexander Sellerholm: {\rm sellerholm@physto.se} \newline 
Gabrijela Zaharijas: {\rm gabrijela.zaharijas@cea.fr}}

\medskip
{\raggedright\small
\begin{center}
A.~A.~Abdo$^{1,2}$, 
M.~Ackermann$^{3}$, 
M.~Ajello$^{3}$, 
L.~Baldini$^{4}$, 
J.~Ballet$^{5}$, 
G.~Barbiellini$^{6,7}$, 
D.~Bastieri$^{8,9}$, 
K.~Bechtol$^{3}$, 
R.~Bellazzini$^{4}$, 
B.~Berenji$^{3}$
% ...
\end{center}

%\newpage
\medskip
%\thispagestyle{empty}
%\begin{enumerate}%\begin{itemize}\addtolength{\itemsep}{-0.5\baselineskip}
   \hspace{3mm}\footnotemark[1] Space Science Division, Naval Research Laboratory, Washington, DC 20375, USA
\\ \hspace{3mm}\footnotemark[2] National Research Council Research Associate, National Academy of Sciences, Washington, DC 20001, USA
\\ \hspace{3mm}\footnotemark[3] W. W. Hansen Experimental Physics Laboratory, Kavli Institute for Particle Astrophysics and Cosmology, Department of Physics and SLAC National Accelerator Laboratory, Stanford University, Stanford, CA 94305, USA
\\ \hspace{3mm}\footnotemark[4] Istituto Nazionale di Fisica Nucleare, Sezione di Pisa, I-56127 Pisa, Italy
\\ \hspace{3mm}\footnotemark[5] Laboratoire AIM, CEA-IRFU/CNRS/Universit\'e Paris Diderot, Service d'Astrophysique, CEA Saclay, 91191 Gif sur Yvette, France
\\ \hspace{3mm}\footnotemark[6] Istituto Nazionale di Fisica Nucleare, Sezione di Trieste, I-34127 Trieste, Italy
\\ \hspace{3mm}\footnotemark[7] Dipartimento di Fisica, Universit\`a di Trieste, I-34127 Trieste, Italy
\\ \hspace{3mm}\footnotemark[8] Istituto Nazionale di Fisica Nucleare, Sezione di Padova, I-35131 Padova, Italy
\\ \hspace{3mm}\footnotemark[9] Dipartimento di Fisica ``G. Galilei", Universit\`a di Padova, I-35131 Padova, Italy
\\ \hspace{3mm}\footnotemark[10] Istituto Nazionale di Fisica Nucleare, Sezione di Perugia, I-06123 Perugia, Italy
\\ \hspace{3mm}\footnotemark[11] Dipartimento di Fisica, Universit\`a degli Studi di Perugia, I-06123 Perugia, Italy
\\ \hspace{3mm}\footnotemark[12] Dipartimento di Fisica ``M. Merlin" dell'Universit\`a e del Politecnico di Bari, I-70126 Bari, Italy
\\ \hspace{3mm}\footnotemark[13] Istituto Nazionale di Fisica Nucleare, Sezione di Bari, 70126 Bari, Italy
\\ \hspace{3mm}\footnotemark[14] Laboratoire Leprince-Ringuet, \'Ecole polytechnique, CNRS/IN2P3, Palaiseau, France
\\ \hspace{3mm}\footnotemark[15] Department of Physics, University of Washington, Seattle, WA 98195-1560, USA
\\ \hspace{3mm}\footnotemark[16] Institut de Ciencies de l'Espai (IEEC-CSIC), Campus UAB, 08193 Barcelona, Spain
\\ \hspace{3mm}\footnotemark[17] INAF-Istituto di Astrofisica Spaziale e Fisica Cosmica, I-20133 Milano, Italy
\\ \hspace{3mm}\footnotemark[18] NASA Goddard Space Flight Center, Greenbelt, MD 20771, USA
\\ \hspace{3mm}\footnotemark[19] Center for Research and Exploration in Space Science and Technology (CRESST) and NASA Goddard Space Flight Center, Greenbelt, MD 20771, USA
\\ \hspace{3mm}\footnotemark[20] Department of Physics and Center for Space Sciences and Technology, University of Maryland Baltimore County, Baltimore, MD 21250, USA
\\ \hspace{3mm}\footnotemark[21] George Mason University, Fairfax, VA 22030, USA
\\ \hspace{3mm}\footnotemark[22] Laboratoire de Physique Th\'eorique et Astroparticules, Universit\'e Montpellier 2, CNRS/IN2P3, Montpellier, France
\\ \hspace{3mm}\footnotemark[23] Department of Physics, Stockholm University, AlbaNova, SE-106 91 Stockholm, Sweden
\\ \hspace{3mm}\footnotemark[24] The Oskar Klein Centre for Cosmoparticle Physics, AlbaNova, SE-106 91 Stockholm, Sweden
\\ \hspace{3mm}\footnotemark[25] Royal Swedish Academy of Sciences Research Fellow, funded by a grant from the K. A. Wallenberg Foundation
\\ \hspace{3mm}\footnotemark[26] Agenzia Spaziale Italiana (ASI) Science Data Center, I-00044 Frascati (Roma), Italy
\\ \hspace{3mm}\footnotemark[27] Dipartimento di Fisica, Universit\`a di Udine and Istituto Nazionale di Fisica Nucleare, Sezione di Trieste, Gruppo Collegato di Udine, I-33100 Udine, Italy
\\ \hspace{3mm}\footnotemark[28] CNRS/IN2P3, Centre d'\'Etudes Nucl\'eaires Bordeaux Gradignan, UMR 5797, Gradignan, 33175, France
\\ \hspace{3mm}\footnotemark[29] Universit\'e de Bordeaux, Centre d'\'Etudes Nucl\'eaires Bordeaux Gradignan, UMR 5797, Gradignan, 33175, France
\\ \hspace{3mm}\footnotemark[30] Osservatorio Astronomico di Trieste, Istituto Nazionale di Astrofisica, I-34143 Trieste, Italy
\\ \hspace{3mm}\footnotemark[31] Department of Physical Sciences, Hiroshima University, Higashi-Hiroshima, Hiroshima 739-8526, Japan
\\ \hspace{3mm}\footnotemark[32] Department of Astronomy and Astrophysics, Pennsylvania State University, University Park, PA 16802, USA
\\ \hspace{3mm}\footnotemark[33] Department of Physics and Department of Astronomy, University of Maryland, College Park, MD 20742, USA
\\ \hspace{3mm}\footnotemark[34] Max-Planck-Institut f\"ur Radioastronomie, Auf dem H\"ugel 69, 53121 Bonn, Germany
\\ \hspace{3mm}\footnotemark[35] Center for Space Plasma and Aeronomic Research (CSPAR), University of Alabama in Huntsville, Huntsville, AL 35899, USA
\\ \hspace{3mm}\footnotemark[36] Instituci\'o Catalana de Recerca i Estudis Avan\c{c}ats (ICREA), Barcelona, Spain
\\ \hspace{3mm}\footnotemark[37] Department of Physics, Center for Cosmology and Astro-Particle Physics, The Ohio State University, Columbus, OH 43210, USA
\\ \hspace{3mm}\footnotemark[38] Research Institute for Science and Engineering, Waseda University, 3-4-1, Okubo, Shinjuku, Tokyo, 169-8555 Japan
\\ \hspace{3mm}\footnotemark[39] Department of Physics, Tokyo Institute of Technology, Meguro City, Tokyo 152-8551, Japan
\\ \hspace{3mm}\footnotemark[40] Cosmic Radiation Laboratory, Institute of Physical and Chemical Research (RIKEN), Wako, Saitama 351-0198, Japan
\\ \hspace{3mm}\footnotemark[41] Centre d'\'Etude Spatiale des Rayonnements, CNRS/UPS, BP 44346, F-30128 Toulouse Cedex 4, France
\\ \hspace{3mm}\footnotemark[42] Istituto Nazionale di Fisica Nucleare, Sezione di Roma ``Tor Vergata", I-00133 Roma, Italy
\\ \hspace{3mm}\footnotemark[43] Department of Physics and Astronomy, University of Denver, Denver, CO 80208, USA
\\ \hspace{3mm}\footnotemark[44] Max-Planck Institut f\"ur extraterrestrische Physik, 85748 Garching, Germany
\\ \hspace{3mm}\footnotemark[45] Institut f\"ur Astro- und Teilchenphysik and Institut f\"ur Theoretische Physik, Leopold-Franzens-Universit\"at Innsbruck, A-6020 Innsbruck, Austria
\\ \hspace{3mm}\footnotemark[46] Santa Cruz Institute for Particle Physics, Department of Physics and Department of Astronomy and Astrophysics, University of California at Santa Cruz, Santa Cruz, CA 95064, USA
\\ \hspace{3mm}\footnotemark[47] Space Sciences Division, NASA Ames Research Center, Moffett Field, CA 94035-1000, USA
\\ \hspace{3mm}\footnotemark[48] NYCB Real-Time Computing Inc., Lattingtown, NY 11560-1025, USA
\\ \hspace{3mm}\footnotemark[49] Department of Chemistry and Physics, Purdue University Calumet, Hammond, IN 46323-2094, USA
\\ \hspace{3mm}\footnotemark[50] Dipartimento di Fisica, Universit\`a di Roma ``Tor Vergata", I-00133 Roma, Italy
\\ \hspace{3mm}\footnotemark[51] Department of Physics, Royal Institute of Technology (KTH), AlbaNova, SE-106 91 Stockholm, Sweden
\\ \hspace{3mm}\footnotemark[52] School of Pure and Applied Natural Sciences, University of Kalmar, SE-391 82 Kalmar, Sweden
\\ \hspace{3mm}\footnotemark[53] Institut de Physique Th{\' e}orique, CEA, IPhT, F-91191 Gif-sur-Yvette, France
%\\ \hspace{3mm}\footnotemark[*] current address
}

\newpage
\bigskip
\centerline{(Dated: February 24, 2010)}
\medskip

%%%%%%%%%%%%%%%%%%%%%%%%%%%%%%%%%%%%%%%%%%%%
\begin{abstract}
The first published Fermi large area telescope (Fermi-LAT) measurement of the isotropic diffuse gamma-ray emission is in good agreement with a single power law, and is not showing any signature of a dominant contribution from dark matter sources in the energy range from 20 to 100 GeV. We use the absolute size and spectral shape of this measured flux to derive cross section limits on three types of generic dark matter candidates: annihilating into quarks, charged leptons and monochromatic photons. Predicted gamma-ray fluxes from annihilating dark matter are strongly affected by the underlying distribution of dark matter, and by using different available results of matter structure formation we assess these uncertainties. We also quantify how the dark matter constraints depend on the assumed conventional backgrounds and on the Universe's transparency to high-energy gamma-rays. In reasonable background and dark matter structure scenarios (but not in all scenarios we consider) it is possible to exclude models proposed to explain the excess of electrons and positrons measured by the Fermi-LAT and PAMELA experiments. Derived limits also start to probe cross sections expected from thermally produced relics (e.g. in minimal supersymmetry models) annihilating predominantly into quarks. For the monochromatic gamma-ray signature, the current measurement constrains only dark matter scenarios with very strong signals.
\end{abstract}

%%%%%%%%%%%%%%%%%%%%%%%%%%%%%%%%%%%%%%%%%%%%
\section{Introduction}
From the early 1980's, when astrophysicists became first largely convinced that most of the mass in the Universe is dark \cite{Peebles:1982ff}, to more recent years when the $\Lambda$CDM paradigm became well established, a substantial experimental effort has been dedicated to dark matter (DM) identification. One strong candidate for the constituents of DM are Weakly Interactive Massive Particles (WIMPs) which, by having masses and couplings at the electroweak scale, naturally produce a relic abundance in agreement with current observations; for a review see, {\it e.g.}, \cite{Jungman:1995df,Bergstrom:2000pn,Bertone:2004pz}. Potentially, one of the methods to indirectly detect DM is its annihilation or decay signal in gamma-rays. In that respect, the data collected by the Fermi-LAT instrument \cite{Atwood:2009ez} have been eagerly awaited. It explores the gamma-ray sky in the 20 MeV to 300 GeV range and has an improved sensitivity by more than an order of magnitude, compared to its predecessor EGRET \cite{egret}. After its first year of mission, it has already provided a wealth of information and advanced our knowledge in many areas of astrophysics and has set new limits on the rate of DM induced gamma-rays \cite{Scott:2009jn,fermidwarfs,fermilines,Cirelli:2009dv,Papucci:2009gd,Pohl:2009qt,fermiclusters}. Some of the promising DM gamma-ray targets are the Galactic center \cite{Serpico:2008ga,Dodelson:2007gd}, because of its proximity and expected high density of DM; dwarf satellite galaxies\cite{Strigari:2006rd,Strigari:2007at}, because of their low or absent astrophysical backgrounds; the extended Milky Way halo \cite{Pieri:2007ir,Springel:2008zz}; and individual DM substructures in our Galaxy \cite{Kuhlen:2009is,Kuhlen:2009kx}.

In this work we use the full sky gamma-ray survey to investigate a possible isotropic DM signal in  originating from annihilations summed over halos at all redshifts \cite{Bergstrom:2001jj,Ullio:2002pj,Taylor:2002zd}. Most cosmological halos are individually unresolved and will contribute to an approximately isotropic gamma-ray background radiation (IGRB). Attempts to fit a cosmological DM signal to the IGRB previously measured by EGRET \cite{Sreekumar:1997un,Strong:2004ry} have been presented in \cite{Elsaesser:2004ap}, where they find that a neutralino, as the lightest supersymmetric particle, with a mass $M_\chi \approx 500 $ GeV gives the best fit to the data. However, due to the large uncertainties both in the data and in the background model they do not claim a detection (see also \cite{deBoer:2007zc,Ibarra:2007wg} for other attempts to fit a DM signal). In the new IGRB measurement presented by the Fermi-LAT in  \cite{egbpaper}, there are no spectral features that favor any of these DM candidates.

Using the Fermi-LAT measured isotropic diffuse gamma-ray background emission \cite{egbpaper}, we present new limits on a signal from cosmological DM. Besides such a signal, the IGRB could also receive a contribution from unresolved Galactic DM subhalos \cite{Diemand:2007qr,Springel:2008cc}. We will focus mainly on limiting the extragalactic DM signal in this work, but comment carefully on the possible size of Galactic contributions. A different approach to extract a DM signal from a full sky analysis, which we will not follow, is to analyze the power spectrum of the gamma-ray signal, which may contain identifiable signatures on different angular scales \cite{Ando:2005xg,Cuoco:2007sh,SiegalGaskins:2008ge,Fornasa:2009qh,SiegalGaskins:2009ux}. 

There are several important uncertainties inherently present when trying to constrain DM properties from the type of analysis presented in \cite{egbpaper}. The largest comes from the theoretical modeling of the expected DM annihilation luminosity. We use recently presented results from the `Millennium II' simulation of cosmic structure formation \cite{Zavala:2009zr,BoylanKolchin:2009nc}, as well as the approach in the Fermi-LAT pre-launch study \cite{Baltz:2008wd}, to calculate the DM contribution to the IGRB signal. Another uncertainty stems from the contribution of more conventional, astrophysical sources to the extragalactic gamma-ray signal, which is currently hard to quantify. A large contribution is believed to originate from unresolved point sources, with the most important potentially being unresolved blazars 
%(in particular the classes of Flat Spectrum Radio Quasars (FSRQs) and BL Lacs) 
\cite{pohl,Giommi:2005bp,Narumoto:2006qg,Dermer:2006pd,Venters:2009sd}. Other sources, such as ordinary star forming galaxies \cite{Pavlidou:2002va,Ando:2009nk} and in particular starburst galaxies \cite{Thompson:2006qd}, as well as structure shocks in clusters of galaxies \cite{Waxman:2000pf,Miniati:2002hs,Keshet:2002sw,Blasi:2007pm,Miniati:2007ke}, might also contribute (see, {\it e.g.}, \cite{Dermer:2007fg} for a short review). The Fermi-LAT is expected to improve our knowledge of these sources and increase our understanding of the shape and normalization of their contribution to the IGRB in the near future (for early results, see \cite{Marco}). We address these background uncertainties by presenting both very conservative and more theoretically-motivated limits on the DM contribution to the IGRB signal.

\bigskip
The paper is organized as follows. In section \ref{flux} we describe the calculation of the isotropic gamma-ray flux from cosmological distant DM annihilations, and comment on the potential contribution from Galactic DM. In section \ref{models} we motivate and describe the particle physics DM models we constrain.  Section \ref{limits} contains a description of our procedure for obtaining the limits, and in section \ref{results} we present and discuss our results. Section \ref{summary} contains our summary.

%%%%%%%%%%%%%%%%%%%%%%%%%%%%%%%%%%%%%%%%%%%%
\section{Dark Matter Induced Isotropic Gamma-Ray Flux} \label{flux}

%%%%%%%%%%%%%%%%%%%%%%%%%%%%%%%%%%%%%%%%%%%%
\subsection{Extragalactic}\label{sec:EG}
\noindent
There are several ingredients necessary to calculate the flux of gamma-rays from cosmological DM annihilation. In addition to the gamma-ray yield per annihilation, assumptions need to be made on the distribution and evolution of DM halos in the Universe. Also, for high-energy gamma-rays, the effects of intergalactic absorption become important and has to be taken into account.
%The dominant contribution to absorption in the tens of GeV to TeV range is pair production on the extragalactic background light emitted by galaxies in the optical and infra-red range.
The flux from DM induced extragalactic photons can be expressed as, \cite{Ullio:2002pj},
\begin{equation}
\frac{d\phi_\gamma}{dE_0} = \frac{\langle\sigma v \rangle}{8 \pi} \frac{c}{H_0} \frac{\bar{\rho}_0^2}{m_{DM}^2} \int{dz (1+z)^3} \frac{\Delta^2(z)}{h(z)} \frac{dN_\gamma(E_0(1+z))}{dE} e^{-\tau(z,E_0)}, \label{eq:1}
\end{equation}
where $c$ is the speed of light, $H_0$ the Hubble constant equal to 100$\times h$%\footnote{not to be confused with $h(z)$}
\,km\,s$^{-1}$/Mpc, $\tau(z,E_0)$ the optical depth, $\langle\sigma v\rangle$ the sample averaged DM annihilation cross section times relative velocity (hereinafter referred to as {\it cross section}), $dN_\gamma/dE$ the gamma-ray spectrum at emission, $m_{DM}$ the DM mass, and $\bar{\rho}_0$ its average density today, while $h(z)=\sqrt{\Omega_M(1+z)^3 + \Omega_\Lambda}$ parameterizes the energy content of the Universe.  
The quantity $\Delta^2(z)$, as defined in \cite{Ullio:2002pj}, describes the enhancement of the annihilation signal arising due to the clustering of DM into halos and subhalos (relative to a uniform DM distribution in the Universe). For the $\Omega_M$, $\Omega_\Lambda$, and $h$ we will consistently adopt the values used in \cite{Ullio:2002pj} and \cite{Zavala:2009zr}; which will be the two references we follow in order to derive $\Delta^2(z)$.
%and is taken from the WMAP three-year data cite{Spergel:2006hy}. 

To quantify $\Delta^2(z)$ it is necessary to know the DM distribution on all length scales. Currently it is best predicted from numerical N-body simulations, which calculate the evolution of the matter distribution from an  {\it ab initio} almost homogeneous distribution of DM in the early Universe. However, for N-body simulations computing resources limit the mass-resolution and the ability to properly model the effects of baryons. This prevents solid predictions for the DM structure on all scales, from cosmological down to the smallest scales of relevance. As the largest contribution to the DM induced extragalactic gamma-ray flux might come from small halos formed in an earlier, denser Universe, it is of importance to at least extrapolate down to mass scales expected for the smallest DM halos. Also, within the larger halos there should exist smaller bound structures that have survived tidal stripping, and that could contribute significantly \cite{Diemand:2007qr,Springel:2008cc}. The minimal DM halo size is limited by the so called free streaming and/or acoustic oscillations, and is typically in the range $10^{-9} - 10^{-4} M_{\odot}$ for WIMP DM (see, {\it e.g.}, \cite{Green:2003un,Green:2005fa,Loeb:2005pm,Profumo:2006bv,Bringmann:2009vf} and references therein). The needed extrapolations are therefore many orders of magnitude, since even simulations concentrated on Milky Way-size halos do not currently reach below subhalo masses of  about $10^{5} M_{\odot}$ at $z=0$ \cite{Springel:2008cc}. We will take two main approaches to calculate the cosmological DM signal, and consider different extrapolations to different smallest halo masses.
%Also, within the halos, there should exist smaller, bound structures that have survived tidal stripping cite{Diemand:2007qr,Springel:2008cc}. Even though the smallest mass scale for WIMP halos are limited by its free streaming length and/or its acoustic oscillations, typically in the range $~10^{-9} - 10^{-4} M_{\odot}$ cite{Bringmann:2009vf}, the extrapolations from simulation are still many orders of magnitude. 
%Although not as massive as the primary halos, the substructure halos arise in higher density environments, which makes them denser than their surrounding parent halos. Two recent projects studying the structure of Galaxy-sized halos cite{Diemand:2007qr,Springel:2008cc} and focusing on the inner regions and the substructure population, have confirmed a substantial subhalo population and revealed presence of substructure within subhalos. It has been shown that even modest levels of substructures can boost the DM EGB signal by approximately an order of magnitude, cite{Ullio:2002pj}.

In the first approach, we will use the results from one of the most recent N-body simulations, `Millennium II' (MS-II) \cite{BoylanKolchin:2009nc}, as it was used by  Zavala {\it et al.}\ in \cite{Zavala:2009zr} to determine the cosmological DM annihilation signal. In their work, two basic structural properties of DM halos, the maximal rotational velocity $V_{max}$ and the radius where the maximum occurs, $r_{max}$, are used to characterize their structure. By approximating the internal structure of halos by spherically symmetric Navarro, Frenk and White (NFW) density profiles \cite{Navarro:1996gj}, $V_{max}$ and $r_{max}$ enable calculation of the luminosity expected from halos down to a fairly low mass limit, $\sim 7~10^8~h^{-1}~M_{\odot}$ ({\it i.e.}, the resolution of resolved halos in the simulation). If instead an Einasto profile shape \cite{einasto,einastoAQ} was assumed, that would only introduce an increase of about 50$\%$ in the signal, as argued in \cite{Zavala:2009zr}. 

To take into account the presence of host halos smaller than the ones resolved in the simulation, Zavala {\it et al.}\ fit a single power law (for each redshift) to the differential luminosity contribution per halo mass interval {\it versus} halo mass. The obtained luminosity function is then extrapolated down to a damping scale mass limit of $10^{-6}  h^{-1} M_\odot$.\footnote{We note that this extrapolation breaks down for redshifts $z\gtrsim 2.1$. However, the dominant contribution to the DM extragalactic signal comes from low redshifts, with the contribution to the signal from redshifts above 2 being at a $\lesssim 30 \%$ level \cite{Profumo:2009uf}. This statement stays true for all DM particle scenarios, DM structure and absorption models we consider.}
%The uncertainty in the power law parameters is what give the main uncertainty in the calculation of $\Delta^2(z)$ down to the 
To also include the effects from substructures inside main halos, a similar universal power law was also independently established for the differential luminosity contribution as a function of subhalo mass. This power law for substructures was obtained by using the most massive halos in the MS-II that still have a significant number of resolved subhalos.  
Again this power law was used to extrapolate the luminosity contribution from subhalos  down to $10^{-6}  h^{-1} M_\odot$. Using reasonable uncertainty bounds, both on normalization and slope in the power law, for the undertaken extrapolation, Zavala {\it et al.}\ find that the enhancement factor due to substructures can be between roughly 2 and 2000 for a Milky-Way-sized halo. This is in broad agreement with the value of 232 presented in \cite{Springel:2008zz} for the high resolution Aq-A halo of the Aquarius project (which is a part of the Millennium simulation suite).

When placing our DM limits, we consider three cases for $\Delta^2$ from the derivation in \cite{Zavala:2009zr}. The most conservative case takes into account the contribution to the DM signal only from halos and subhalos resolved in MS-II simulation (MSII-Res).  As the other extreme, we consider the most optimistic case, where subhalos and halos down to $\sim 10^{-6}M_{\odot}$ was taken into account as to maximize the contribution of these smallest halos (MSII-Sub2).  We also consider a moderate case (MSII-Sub1), which we will adopt as our reference model, in which the extrapolation of the contribution from structures and substructures down to $\sim10^{-6}M_{\odot}$ has been done in a very conservative way.

\smallskip
As an alternative approach, we calculate $\Delta^2$ using the same `semi-analytical' procedure as in \cite{Ullio:2002pj}. Here the contribution from halos of all masses is integrated by using theoretically motivated analytical functional forms on the relevant properties of the DM structure, which in turn are tuned to fit results from numerical N-body simulations. The ingredients needed are the redshift dependence on the halo mass function, the DM halo density profile, and the spread in halo shapes for each halo mass.  It is convenient to parametrize a halo by its virial mass, $M$, and concentration parameter, $c$. 
%The concentration parameter is defined as $c = R/r_{-2}$, where $r_ {-2}$ is the distance where the profile falls as $r^{-2}$ and $R$ is the virial radius ....
The concentration parameter is treated as a stochastic variable with a log-normal distribution for a fixed mass. Simplified, we can write the quantity $\Delta^2(z)$ as:
\begin{equation}
\Delta^2(z)=\int\,dM\,\frac{dn}{dM}\int\,dc\,P(c)\frac{\langle \rho^2(M,\,c)
 \rangle }{\langle \rho(M,\,c)\rangle ^2}
\end{equation}
where $dn/dM$ is the halo mass function, with its functional form calculated as in the ellipsoidal collapse model \cite{Sheth:1999su}, %and fit to the N body simulation of Virgo Consortium cite{Jenkins:1997en}} 
and $P(c)$ is a log-normal distribution with variance $\sigma(\log_{10}c)=0.2$ \cite{Bullock:1999he,Ullio:2002pj}. We model the dependence of concentration parameter on halo mass and redshift according to the Bullock {\it et al}.\ toy model \cite{Bullock:1999he}.  We assume a NFW DM profile and fix the concentration parameter to stay constant for halo masses below $10^5 M_\odot$, as it was done in \cite{Ullio:2002pj}, to minimize the risk of overestimating the DM signal from extrapolation of the model of Bullock {\it et al.}\ to low halo masses. Similarly to \cite{Ullio:2002pj,Baltz:2008wd}, we set $10 \%$ of a halo mass in substructures and assume that the subhalo mass function has a power-law behavior in mass $M^{-\beta}$, with a slope $\beta =1.9$. This is in broad agreement with findings of new simulations \cite{Diemand:2007qr,Springel:2008cc} for Milky Way-size halos.  The concentration parameter of subhalos is not constant, but depends on the subhalo mass \cite{Ullio:2002pj} and on the distance from the center of the halos \cite{Diemand:2007qr,Springel:2008cc}. We here associate a concentration parameter four times higher in substructures, compared to a main halo of the same mass \cite{Ullio:2002pj}. This is the same type of structure description also used in the Fermi pre-launch paper \cite{Baltz:2008wd}, and used in several recent works \cite{Profumo:2009uf,Belikov:2009cx}. We dub this scenario the semi-analytical NFW Bullock {\it et al.}~substructure model (BulSub). 

\begin{figure}[t]
\centerline{\includegraphics[width=0.95\columnwidth]{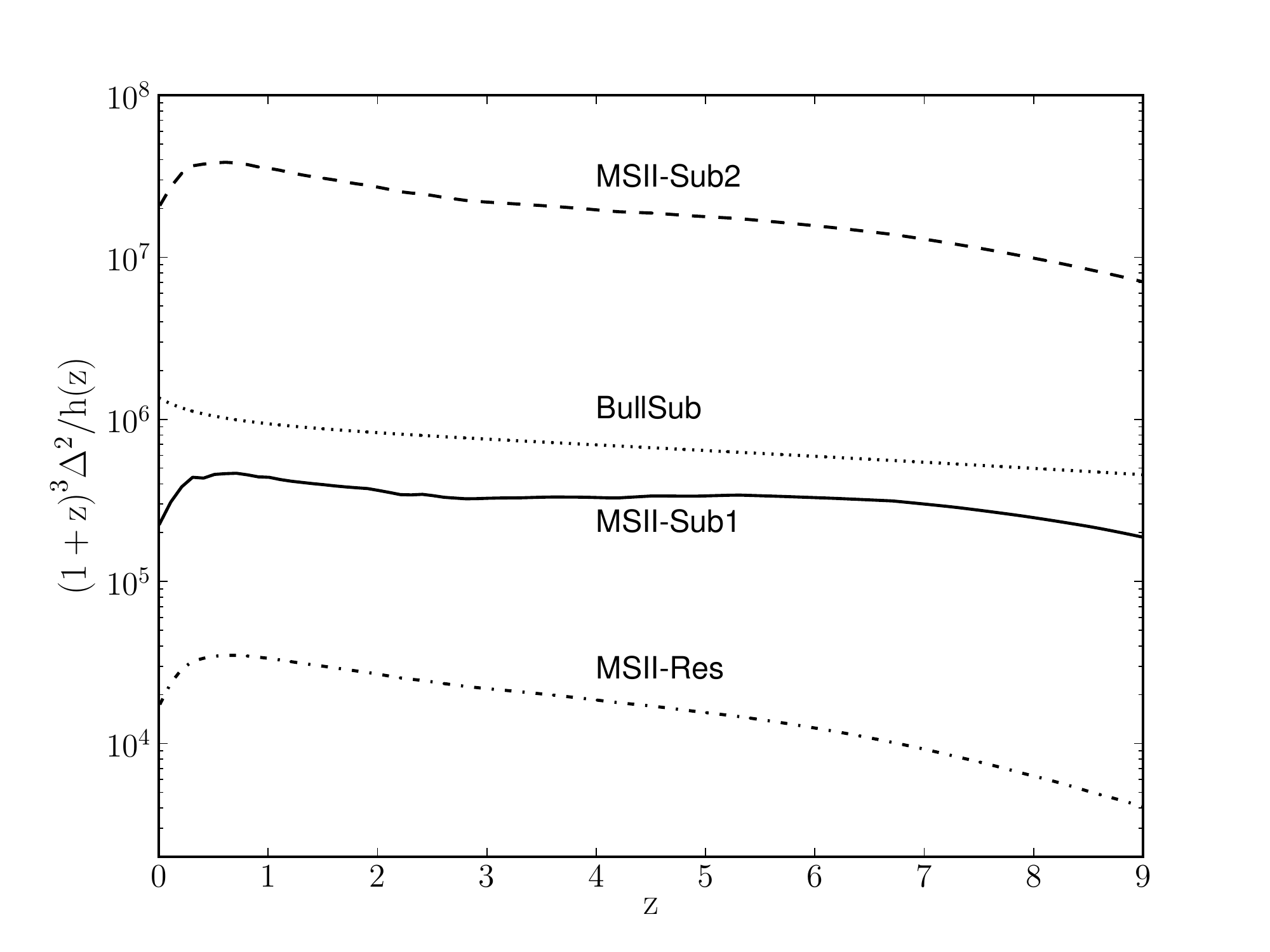}}
\caption[]{Comparison of the different models used to calculate the enhancement of DM annihilation signal due to structure formation; $\Delta^2 (z)$ based on the Millennium II simulation (MSII-models) \cite{Zavala:2009zr} and the semi-analytic model (BulSub) \cite{Ullio:2002pj}.  All the enhancement factors $\Delta^2 (z)$ are multiplied by the factor $(1+z)^3/h(z)$ in order to reflect the relevant part of the integrand in equation (\ref{eq:1}) we want to illustrate.} \label{delta}
\end{figure}	

The result of the semi-analytical (BulSub) approach lies between the extreme values found in MS-II simulation, and turns out to be quite close to the MSII-Sub1 case. We show a comparison of these four models via the quantity $(1+z)^3 \Delta^2(z)/h(z)$ in figure~\ref{delta}. The difference in shape, at low redshifts, between the semi-analytical (BulSub) model and the MS-II results comes mainly from different redshift evolution of the concentrations parameter and halo mass function. To summarize, we note that all above scenarios could basically be related by an overall shift in their predicted signal amplitudes. We will keep all of them, however, in our exclusion plots as it is illustrative and allows easy comparisons with previous works.

\begin{figure}[t]
\centerline{\includegraphics[width=0.95\columnwidth]{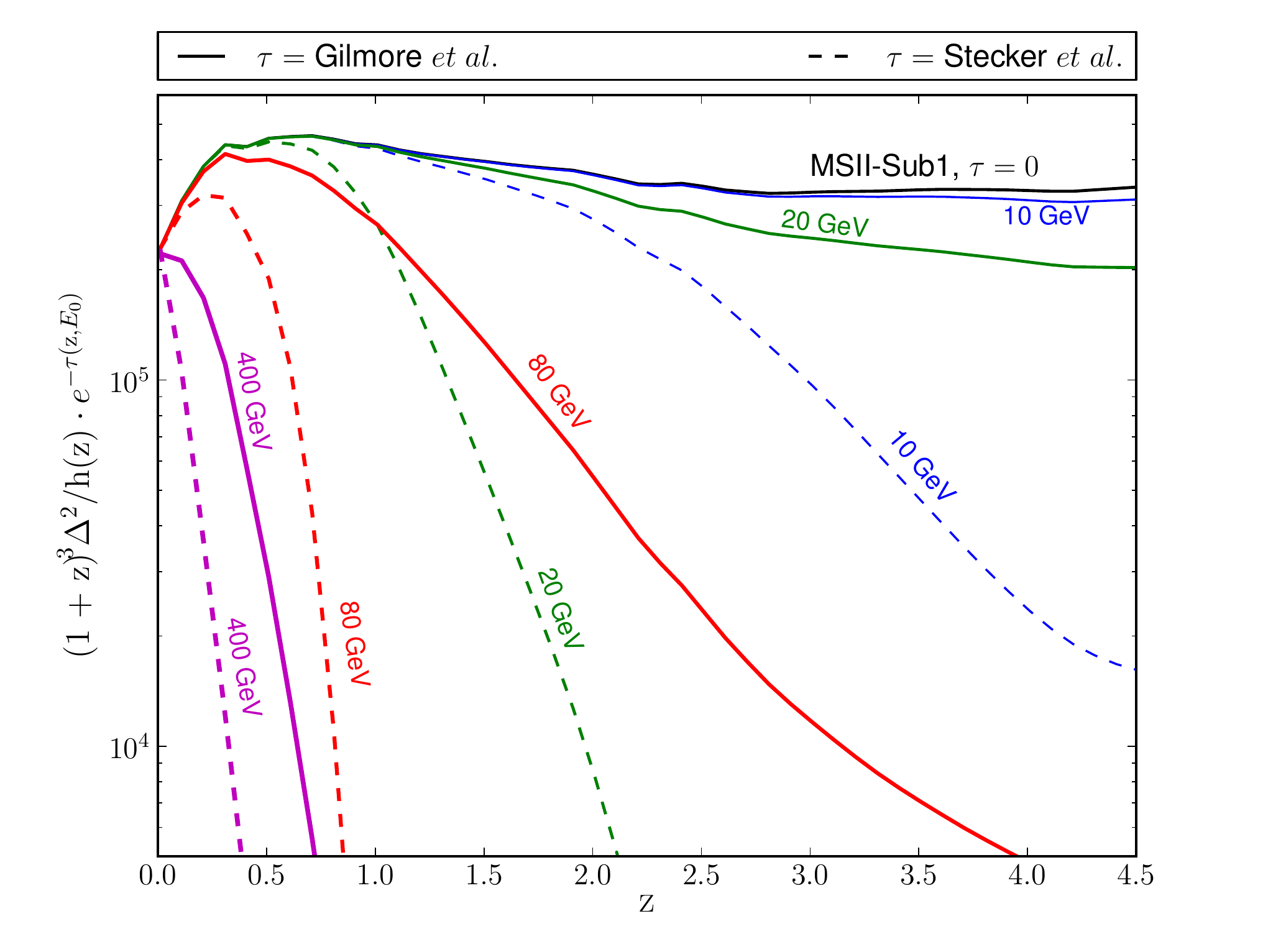}}
   \caption{Comparison of the gamma-ray absorption models of Gilmore {\it et al.}\ \cite{Gilmore:2009zb} (solid) and Stecker {\it et al.}\ \cite{Stecker:2008fp} (dashed), and their affect on the signal in the MSII-Sub1 structure formation scenario.  The upper most (black solid) line is if no absorption is present.}
   \label{delta_abs}
\end{figure}

\medskip
For the optical depth $\tau(z,E_0)$, as a function of redshift $z$ and observed energy $E_0$, we use the result of Gilmore {\it et al.}\ \cite{Gilmore:2009zb}.\footnote{We implemented the fiducial 1.2 model from \cite{Gilmore:2009zb}.} In figure \ref{delta_abs}, we compare this result to the older, commonly assumed, absorption model of Stecker {\it et al.}\ \cite{Stecker:2008fp} by plotting the relevant part of the integrand in equation (\ref{eq:1}): $(1+z)^3 \Delta^2(z)/h(z) {\rm exp}(-\tau)$. The newer absorption model in \cite{Gilmore:2009zb} gives significantly lower optical depth. For $z\geq1$ the difference to the older model \cite{Stecker:2008fp} is large for gamma-ray energies $E_{0}\gtrsim 20$ GeV, and for higher energies the difference is even larger and their deviation start at much lower redshifts. We show that the choice of absorption model plays a role for the DM limits when the limits are set by the gamma-ray spectrum in the high energy end of the Fermi-LAT measurement. We comment further on this in sections \ref{models} and \ref{results}.

%%%%%%%%%%%%%%%%%%%%%%%%%%%%%%%%%%%%%%%%%%%%
\subsection{Galactic}
In addition to an extragalactic DM signal, there could be a significant contribution from pair annihilations along the line of sight through the DM halo in which the Milky Way is embedded. Current N-body simulations show highly galactocentric smooth DM density profiles, extending far beyond the visible Galaxy, and with the main halo hosting a large amount of substructures in form of subhalos (which themselves contain subhalos) \cite{Springel:2008zz,Diemand:2007qr}. 

The Galactic main halo's DM density profile would by itself, from an observer on Earth, give rise to a very anisotropic DM annihilation signal.\footnote{In \cite{Ando:2005hr} it was also argued that without, {\it e.g.}, a substructure signal enhancements, the observation of the inner degrees of the Milky Way is typically expected to always reveal a DM signal prior to a observed DM gamma-ray signature in the IGRB measurment.}
The DM annihilation signal from the Galactic substructures, however, has a completely different morphology and could potentially produce a fully isotropic signal. This is because the flux is proportional to the number density distribution of subhalos, and this distribution is much less centrally concentrated than the main halo DM distribution %smooth DM annihilation signal 
\cite{Springel:2008cc}. How isotropic the emission from unresolved Galactic subhalos is may vary somewhat between different high resolution N-body simulations (see, {\it e.g.}, \cite{Pieri:2009je}). With a substructure signal spatially distributed as presented in  \cite{Springel:2008zz} for the Aquarius A-1 simulation, we find that the substructure component itself gives a diffuse emission that varies by less than 25 \% in different sky directions. 
Since variation in the substructure distribution between different DM halo realizations is not excluded, we conclude that the signal from unresolved Galactic substructures could be practically isotropic \cite{Springel:2008zz}.

\begin{figure}[t]
\centerline{\includegraphics[width=0.95\columnwidth]{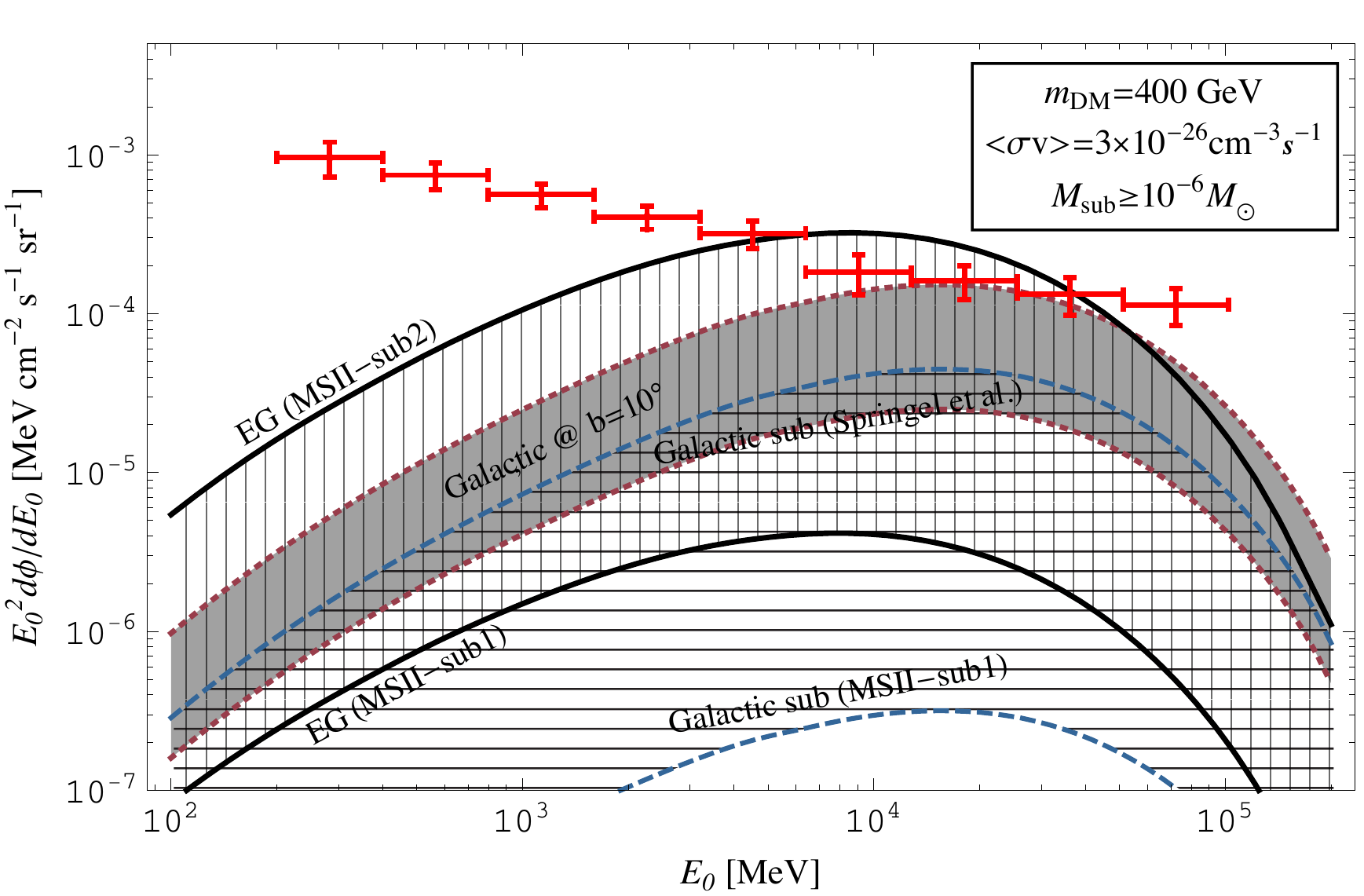}}
   \caption{ The vertically hatched band illustrates the span in the expected isotropic extragalactic (EG) gamma-ray signal, defined by being the region enclosed by our MSII-Sub1 and MSII-Sub2 cases. The horizontally hatched band is the flux that can be expected from Galactic substructure. The filled grey band is the signal range that could be expected from the main DM Galactic halo, at a latitude of 10$^\circ$, which would by itself produce an anisotropic signal. The data points show the measurement of the IGRB by the Fermi-LAT \cite{egbpaper}.  The gamma-ray spectra are from DM particles with mass of 400 GeV, a total annihilation cross section  $\langle \sigma v\rangle= 3\times 10^{-26}$ cm$^{3}$ s$^{-1}$ into  $b {\bar b}$ quarks, and a minimal subhalo mass cut-off at $10^{-6} M_{\odot}$.  See the text for more details.}
   \label{Gal}
\end{figure}

To illustrate the range of predictions for the isotropy of the DM-induced gamma-ray spectra, we show in figure \ref{Gal} uncertainty bands for the three mentioned DM components ({\it i.e.}\ Galactic main halo, Galactic substructure, and the extragalactic component), for DM annihilation into $b {\bar b}$ quarks.

The vertically hatched band illustrates the span in the expected (isotropic) extragalactic signal, defined as the region between our MSII-Sub1 and MSII-Sub2 cases. 

The horizontally hatched band is the ($\sim$ isotropic) flux that can be expected from Galactic DM substructure. The band's width indicates the range in signal strength, and comes from uncertainties in how the extrapolation of the flux contribution from DM subhalos down to masses of $10^{-6} M_{\odot}$ is performed. We model the gamma-ray signal from Galactic substructures, inside the galactocentric distance $r$ (in kpc), as:
\begin{equation} 
\mathcal L_{\rm sub} (<r) = \mathcal L^{200}_{\rm main} \times B \times x^{x^{-0.24}},  \;\;\hbox{where $x=r/r_{200}$ and $r_{200}  \approx 200$ kpc.}
\end{equation}
This functional form is a parametrization of the result presented for the Aquarius simulation in \cite{Springel:2008zz}. $\mathcal L^{200}_{\rm main}$ is the total DM-induced luminosity inside $r_{200}$ from the smooth halo (normalized through the Einasto profile in equation (\ref{eq:profile})),  and $B$ gives the relative signal enhancement inside $r_{200}$ due to substructures.
The upper boarder of the vertically hatched band is obtained when a single power law relation between the substructure flux and the minimal DM subhalo mass are related as suggested in  \cite{Springel:2008zz}, which give $B \sim 230$.\footnote{We note that by using the MSII-Sub2 prescription for substructure for Milky Way sized halos, the vertically hatched upper limit would be extended up further by one order of magnitude.}  
The lower boarder is when the substructure signal strength instead is implemented consistently with the average  substructure enhancement used in the MSII-Sub1 calculation of the extragalactic signal. Then the luminosity from all substructures inside $r_{200}$ for a Milky-Way-sized halos is merely $B\sim 2$ times the luminosity of the main DM halo. This lower signal limit is also similar in amplitude to the finding in \cite{Pieri:2009je}, where the Aquarius simulation is used, but a subhalo concentration extrapolation with a double power law approach is applied to soften the DM halo concentration for small subhalo masses. 
We thus find that the diffuse DM signal from Galactic substructure could be insignificant, but that, with the uncertainty bands in figure~\ref{Gal}, Galactic substructures could also potentially enhance the DM signal by at least an order of magnitude relative to the extragalactic MSII-Sub1 signal. This range covers the result that  \cite{Pieri:2009je} finds by self-consistently extrapolating results from two specific high resolution simulated halos. All these scenarios would obviously only increase the DM signal and would, if taken into account, only lead to stronger DM constraints than we derive from the  extragalactic signal in this work.

The grey band is the (anisotropic) signal from the main DM Galactic halo, at a latitude of 10$^\circ$, where the upper and lower limits correspond to when an Einasto and a cored isothermal \cite{ISO:Bahcall:1980fb} DM density profile is adopted, respectively.   These profiles are given by:
\begin{equation} 
\frac{\rho(r)}{\rho_\odot} 
= \left\{\begin{array}{ll}
(1+r_\odot^2/r_s^2)/(1+r^2/r_s^2) & \hbox{isothermal, $r_s = 5\,{\rm kpc}$}\\
\exp(-2[(r/r_s)^\alpha - (r_\odot/r_s)^\alpha]/\alpha)& \hbox{Einasto, $r_s = 20\,{\rm kpc}$, $\alpha=0.17,$}\\
\end{array}
\right.\label{eq:profile}
\end{equation}
where the local DM density $\rho(r=r_\odot\approx 8.2\,{\rm kpc}) = \rho_\odot=0.4\ {\rm GeV} / {\rm cm}^3$.

\smallskip
For points in the vertically and horizontally hatched bands below the grey band, the anisotropic DM signal from the smooth main halo is expected to become stronger than the isotropic signal (at a latitude of 10$^\circ$).\footnote{In the case of strong `Sommerfeld enhancement' the lower velocity dispersion of smaller objects could cause a relative enhancement in the signal from substructure (or small halos in general) compared to the smooth main Milky Way DM halo. Typically this is a small effect, but in the case of resonances the relative enhancement could be up to a factor $\sim 1000$  \cite{Yuan:2009bb,Kistler:2009xf,Kamionkowski:2010mi},  and correspondingly make the DM signal more isotropic.} 
%\footnote{In cite{Ando:2005hr} it was pointed out that without {\it e.g.} a substructure signal enhancements, the observation of the inner degrees of the Milky Way is typically expected to always reveal a DM signal prior to a observed DM gamma-ray signature in the IGRB. } 

%
 In such scenarios, in principle a different analysis approach should be chosen as a DM component should be included in the modeling of the Galactic (anisotropic) contribution for the IGRB measurement (which was not the case in the analysis in \cite{egbpaper}). Such a combined spectral and morphological DM analysis is beyond the scope of this work, and instead we stay focused on the absolute size and spectral shape of the measured IGRB flux to set the limits on DM signals (see \cite{Dodelson:2009ih}, for additional discussion in this direction). 
%
%{\bf REMOVE/KEEP? 
One could still be tempted to use the fact that, with the utilized Galactic diffuse model, the derived IGRB spectrum is observed to be isotropic  within $\sim 10 \%$  outside of the Galactic plane ($|b|\geq 10^\circ$) \cite{egbpaper}. This means that if an anisotropic DM annihilation signal from the Galactic main DM halo, in the region just above the Galactic plane, is sufficiently strong then this anisotropic signal could be more constrained by the found level of anisotropy than the absolute flux in the studied region.   
However, substantial uncertainties in the modeling of the Galactic diffuse contribution are present, and in the end a simultaneous all sky fit, with both a DM component and a model for the Galactic cosmic rays, gas, and interstellar radiation field, would have to be performed before fully reliable quantitative statements can be made.
%}
Additionally, as discussed above, the relation between the extragalactic and Galactic DM signal is not unique, and this uncertainty makes it important to also establish independent limits on isotropic extragalactic DM signals from the IGRB measurement.

\smallskip
We end this discussion on the Galactic DM signal by concluding
%???? 382: footnote{For similar conclusions see cite{Ando:2005hr}.}
that, to have a isotropic signal, one of the following scenarios should be valid: there exists a strong extragalactic DM signal, {\it e.g.}\ due to substructure enhancements like in the MSII-Sub2 case, or the Galactic substructures themselves give a very strong isotropic flux contribution, or that the DM signal from the inner Milky Way region is relatively low, for example by having a several kpc core in the smooth DM density profile. In the remainder of this work we will derive limits on the extragalactic signal, assuming it dominates the IGRB.

%%%%%%%%%%%%%%%%%%%%%%%%%%%%%%%%%%%%%%%%%%%%
\section{Particle Physics Models} \label{models}
A variety of different extensions of the Standard model of particle physics could in principle produce strong enough fluxes of gamma-rays to be observed by the Fermi-LAT \cite{Baltz:2008wd}. We consider three generic types of DM models with distinctively different gamma-ray signatures, exemplified in figure \ref{spec_fig}:

\begin{figure}[t]
\centerline{\includegraphics[width=0.95\columnwidth,height=0.67\columnwidth]{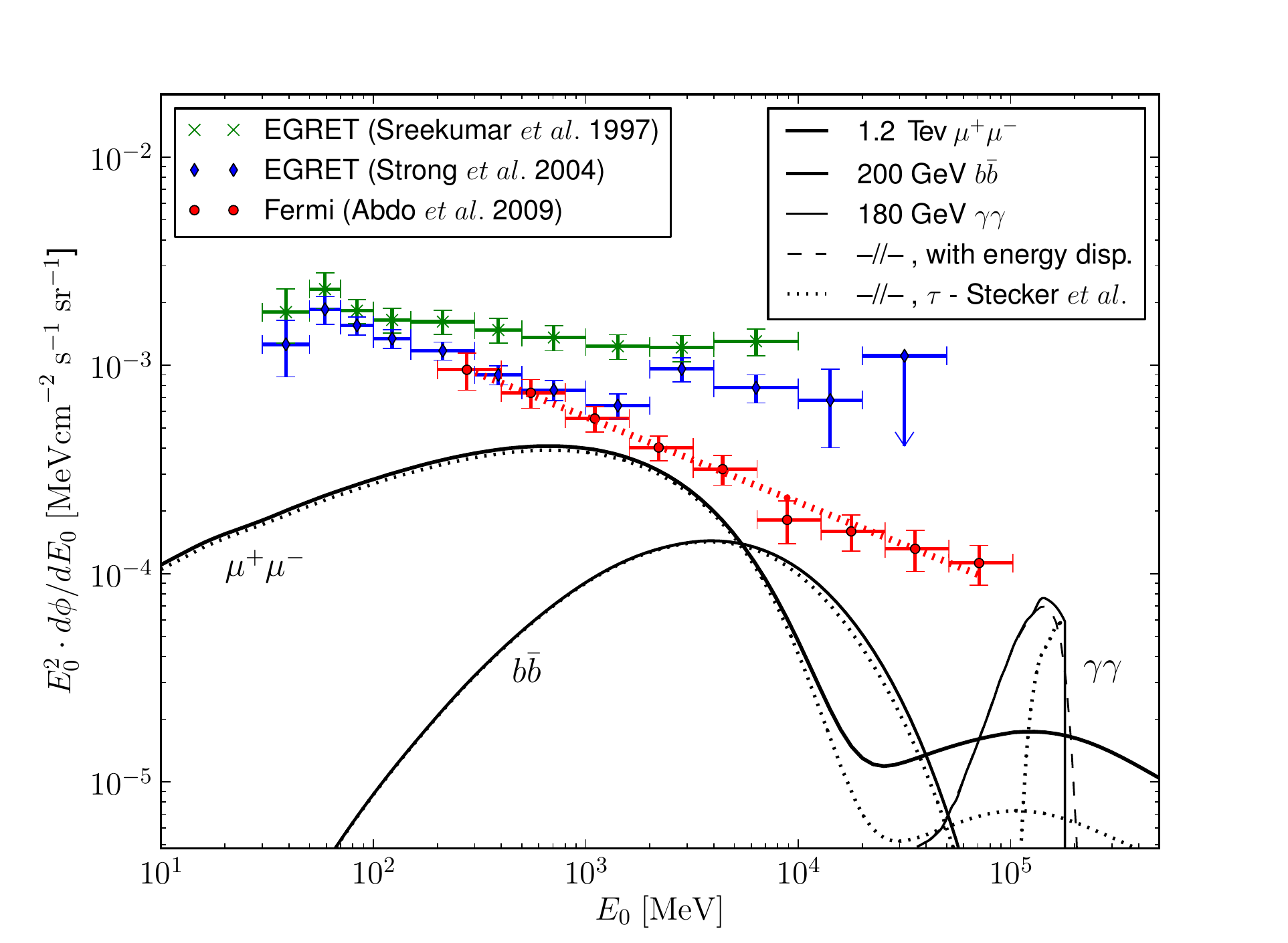}}
   \caption{Measurements of the IGRB by Fermi-LAT \cite{egbpaper} and EGRET \cite{Sreekumar:1997un,Strong:2004ry}, together with three types of gamma-ray spectra induced by DM. The overall normalization of the DM spectra are given by assuming the MSII-Sub1 $\Delta ^2$ model, and for this visualization we have chosen the  following cross sections $\langle \sigma v\rangle= 5\times 10^{-25}$ cm$^{3}$ s$^{-1}$ (for $b {\bar b}$), $1.2\times 10^{-23}$ cm$^{3}$ s$^{-1}$ ($\mu \mu$) and $2.5\times 10^{-26}$ cm$^{3}$ s$^{-1}$ ($\gamma \gamma$). The solid lines are with the Gilmore {\it et al.}\ \cite{Gilmore:2009zb} absorption model applied, and the dotted lines  with the Stecker {\it et al.}\ \cite{Stecker:2008fp} absorption. We also show the line spectra convoluted with the energy resolution of the Fermi-LAT experiment (dashed line).  The dotted line passing through the Fermi data points is a power law with the spectral index of -2.41.}
   \label{spec_fig}
\end{figure}

\begin{itemize}
 \item[]\hspace{-8mm}[Model~1:\ $b \bar{b}$] Many DM candidates ({\it e.g.}, within supersymmetry) have their dominant annihilation channel into quarks and gauge bosons. In these cases, gamma-rays are produced through the hadronization and decay of $\pi^0$. For definiteness we assume here a $100\%$ branching ratio into $b\bar{b}$,  but also annihilations instead into other quarks as well as into W/Z gauge bosons and higgs would all give fairly similar spectra \cite{Hooper:2004vp}.  We use the DarkSUSY package \cite{Gondolo:2004sc} to obtain the gamma-ray yield $dN_\gamma/dE$, when calculating the DM signal. The gammas from inverse Compton (IC) scattering are not included in this case, as they are significantly less constraining.

\item[]\hspace{-8mm}[Model~2:~$\mu^+ \mu^-$]  We consider here DM models having leptons as the dominant annihilation channel. Although  this is not the classical WIMP scenario, these types of models have been suggested to fit the PAMELA positron \cite{Adriani:2008zr} and/or Fermi-LAT electron+positron \cite{Abdo:2009zk} data, without violating anti-proton constraints \cite{Cirelli:2008pk,Donato:2008jk}. The mechanism most often discussed is to either require direct annihilation into leptonic channels {\it ad hoc}, or include a light enough scalar/pseudoscalar which subsequently decays exclusively into leptons due to kinematic constraints (like for example in \cite{Nomura:2008ru,ArkaniHamed:2008qn}). %add cite{Pospelov:2008jd}
These types of (`leptonic') models produce continuous gamma-ray spectra by both final state radiation (FSR) \cite{Bergstrom:2004cy} and inverse Compton scattering \cite{Cholis:2008wq} on background radiation. The annihilation cross section needed is typically higher, by two to three orders of magnitude, than the one naively expected at freeze-out for thermally produced DM. A possibility to obtain such an enhanced signal today could {\it e.g.}\ be by the Sommerfeld enhancement effect \cite{Sommerfeld,Hisano:2003ec,Cirelli:2007xd,ArkaniHamed:2008qn}. We consider the case of $100 \%$ annihilation into mono-energetic $\mu^+$ and $\mu^-$, which is a model that under favorable circumstances has been shown to give a very good fit to cosmic charged lepton data \cite{Bergstrom:2009fa}. 

The yield of gamma-rays, $dN_\gamma/dE$, produced by inverse Compton scattering of DM induced electron and positrons, from the decaying $\mu^+\mu^-$ pairs, off of cosmic background radiation was calculated in \cite{Profumo:2009uf,Belikov:2009cx}. We follow the same approach, using our $\Delta^2(z)$ DM structure scenarios. We convolve the differential Klein-Nishina cross section with the number density of cosmic microwave background radiation %(and interstellar radiation field, which we neglect here) 
and fold it with the spectral distribution of electrons and positrons given by the {\sc DarkSUSY} package \cite{Gondolo:2004sc}, that utilize tabulated {\sc PYTHIA} 6.154 results \cite{Sjostrand:2000wi}. 

The flux of final state radiation photons from the prompt $\mu^+\mu^-\gamma$ process is calculated by using the approximate formula given in \cite{Bergstrom:2004cy,Birkedal:2005ep}. 

Leptonic models favored by the PAMELA and Fermi measurements have been ruled out based on their contribution to radio and gamma-ray signals from the Galactic center region once either an NFW or Einasto DM density profile is chosen \cite{Bergstrom:2008ag,Bertone:2008xr,Regis:2008ij,Cirelli:2009dv,Papucci:2009gd,Crocker:2010gy}. That might seem in contradiction with our approach here, where we, nevertheless, assume a NFW profile for the DM halo distributions. We note, however, that if a cored Burkert profile \cite{Burkert:1995yz} is taken for all halos, the resulting strength of the extragalactic DM signal is only about 20 percent lower compared to the value obtained with a NFW profile \cite{Ullio:2002pj}. On the other hand, the radio and gamma-ray constraints are derived from observation of the inner $\lesssim  {\cal O}(1)$ and $\lesssim  {\cal O}(100)$ parsec, respectively,  from the Galactic center, and a significantly cored DM density profile would avoid these limits  \cite{Bertone:2008xr,Regis:2008ij}.  These inner sub-kiloparsec regions are not yet well resolved by current DM  N-body simulations, and since also the interaction between DM and baryonic components could play an important role there, there is some freedom in tuning the inner DM density in the Milky Way.

 \item[]\hspace{-8mm}[Model~3:~$\gamma \gamma$] In addition to the continuous gamma-spectra, electromagnetically neutral DM can also annihilate through loop suppressed processes and produce monochromatic gamma-lines, either through the annihilation into two photons ($\gamma\gamma$),  a photon and Z-boson ($\gamma Z$)  or a photon and a Higgs-boson ($\gamma h$). While branching ratios to these channels are typically less than $10^{-2}$, there are DM candidates, like the inert doublet model \cite{Gustafsson:2007pc}, non-thermal WIMPs \cite{Kane:2009if} and Dirac DM models with a Z' portal only to heavy quark states in the Standard Model \cite{Jackson:2009kg}, that can have this as their main annihilation channel. For brevity, we consider only the $\gamma \gamma$ channel in this work. In contrast to the continuous photon-spectra for [Model~1] and [Model~2] that have very broad `peaks' at energies of a around a few percent and at around $(m_{\mathrm{DM}}/1 \mathrm{TeV})~10^{-3}$ of the DM particle mass, respectively, the gamma-line sharply peaks at an energy equal to the DM particle mass in the local rest-frame.  Therefore, new limits on this signal are truly enabled by Fermi-LAT measurements, which for the first time provides convincing constraints for gamma-lines in the 10 to 100 GeV energy range. See also the dedicated Fermi-LAT search for Galactic DM lines \cite{fermilines}.

The detection of a multi-GeV gamma line would be an unmistakable DM signal, as no conventional astrophysical source would produce spectral lines at these energies, and give a unique determination of the DM particle mass. However, the observed width of the gamma-line produced in cosmological DM annihilation depends strongly on the optical depth, as illustrated in figure \ref{spec_fig} for the Gilmore {\it et al.}\ \cite{Gilmore:2009zb} and Stecker {\it et al.}\ \cite{Stecker:2008fp} cases. We will illustrate this effect further by calculating DM cross section limits with both choices, in Section \ref{results}. 

The measurement of the IGRB in \cite{egbpaper} is not corrected for the energy dispersion of the instrument. In principle, our signal therefore needs to be convolved with the energy dispersion before comparing it to the data. It can be anticipated that the effect is very small, since the bin width of the measurement is much larger than the energy resolution.  In figure \ref{spec_fig} we convolve the signal from a 180 GeV DM particle with an energy dispersion that follows closely the one used in \cite{fermilines}, which is also valid for the IGRB measurement. The effect on the exclusion limits is, as expected, negligible.
%In figure \ref{spec_fig} we convolve the signal from a 180 GeV DM particle with a Gaussian energy dispersion of 10\% that follows close enough what is valid for the IGRB measurement. 
 
 \end{itemize}

%%%%%%%%%%%%%%%%%%%%%%%%%%%%%%%%%%%%%%%%%%%%
\section{Upper Limit Analysis} \label{limits}
The Fermi IGRB measurement is obtained as an isotropic component in a full sky fit (for latitudes $\geq 10^\circ$) including multiple diffuse Galactic foregrounds and point sources. To reduce the number of misinterpreted charged particles in the IGRB analysis a more stringent event selection (compared to the current public Fermi Pass6V3 diffuse 
%P6_V3, P6\_V3\_DIFFUSE,  http://fermi.gsfc.nasa.gov/ssc/data/analysis/documentation/Cicerone/Cicerone_Data_Exploration/Data_preparation.html
selection class \cite{Atwood:2009ez,nasa}) was imposed. It was estimated that this selection enabled to further reduce the cosmic-ray contamination by 25 - 95\% in the energy range from 0.2 to 102.4 GeV, making the charged cosmic-ray particles subdominant to the IGRB flux. The remaining charged cosmic-ray background was estimated by a Monte Carlo simulation (the result is shown in figure 3 of \cite{egbpaper}) and subtracted from the isotropic signal to obtain the IGRB.  The final IGRB measurement is dominated by systematic uncertainties related to the above-mentioned subtractions as well as by the uncertainties in the effective area of the instrument. For this analysis these systematic errors were added in quadrature to the statistical errors to give the total one standard deviation error. The measured fluxes $M_i$ and bin errors $\Sigma_i$ are taken from the first line of Table~I in \cite{egbpaper}.
There are also other systematic uncertainties attached to the Fermi-LAT measurement of the IGRB not included in the above errors; the most notable is the uncertainties in the diffuse Galactic foreground modeling. It is difficult to accurately incorporate such systematic uncertainties to the limits reported here, especially as the Galactic foreground model has highly nonlinear dependencies on its parameters. This point merits a dedicated analysis which is beyond the scope of this work.

The measured IGRB energy spectrum is in a good agreement\footnote{A single power law with free normalization and spectral index gives a reduced $\chi^2 = 1.29/(9-2) = 0.18$. From a statistical point of view, a reduced $\chi^2$ value much less than 1 is not expected and could indicate  that the estimated uncertainties used are too large. Probably it is an effect of too conservative estimates of the included systematical errors (that is dominating the errors) and that the data points are treated as uncorrelated. We want  to stay conservative at this stage, and keep in mind that the extra uncertainties from the Galactic diffuse foreground modeling is not folded in, and therefore hold on to these potentially too conservative error estimates.} 
with a single power law with a photon spectral index of -2.41. A single dominant source with a significantly different spectral shape can therefore be excluded in this energy range. In the light of this we derive upper limits on DM annihilation rates $\langle\sigma v\rangle$ for our set of generic DM scenarios. Somewhat different limits can be derived depending on what is assumed for the background to the DM signal. To bracket this we derive limits in two ways:

\bigskip
\noindent~1.  [Conservative]  A conservative upper limit on $\langle\sigma v\rangle$  (for a given cosmological DM distribution and for a given particle DM candidate) is placed by restricting DM signals to not exceed the measured intensity in any individual bin by more than some given significance. Of course, this means that to explain all the Fermi IGRB data bins a fully adjustable additional background is needed to explain the gamma-ray flux also in all the energy-bins to which the DM is not contributing enough. The limit for this conservative approach is hence taken to be the lowest $\langle\sigma v\rangle$ for which the integrated DM flux, $\phi^{\gamma}_i$, in any bin equals the measured flux, $M_i$, plus $n$ times the bin error, $\Sigma$. 
In other `words',  we take the upper limit to be $\sup\{ \langle\sigma v\rangle: \phi^{\gamma}_{i} \leq M_{i}+n \times \Sigma_{i} \}$.  
%Our  90, 95 and 99.999\% upper confidence regions are  when $n =1.28,1.64$ and 4.3, respectively. 
Our upper limits, corresponding to 90, 95 and 99.999\% confidence level, are when $n =1.28,1.64$ and 4.3. All our upper limits are derived under the approximation that the probability distributions for the flux in each energy bin are independent and take Gaussian functional forms. The values of $n$ simply correspond to what is needed to retrieve stated confidence levels for one-sided (upper) limit on the $\langle\sigma v\rangle$ parameter (after marginalizing over all other fitting parameters; as in the case of our stringent limits that include additional background parameters).

\bigskip
\noindent~2.  [Stringent] A more realistic scenario is that there is a significant astrophysical background present to which a small DM signal might be superimposed. The exact shape and normalization of such a background is to large extent still unknown. For the purpose of this analysis we use the sum of two power laws as the background. One of the power-law spectral components is motivated by theoretical calculations of the contribution of starforming galaxies to the extragalactic background
\cite{Ando:2009nk}, with a photon spectral index of -2.7. The other one is motivated by the observed spectra of blazars and has a harder photon spectral index of -2.4 \cite{Venters:2009sd,Marco}.\footnote{Note that the astrophysical extragalactic signal is not necessarily expected to be a single power law, or a sum of two. Different sources are expected to have different spectral indices, and a sum of  different power laws is not a single power law. Another consideration is that gamma-gamma absorption against the extragalactic background light would affect the spectral shape at the high end of the LAT energy range \cite{Venters:2009sd}. However, this is a convenient approximation for the present purpose, and the assumption of two power laws enables an excellent fit to data (reduced $\chi^2 = 1.24/(9-2) = 0.18$).}  How much a sum of two such power laws can be distorted by a DM signal, before deviating by more than some given amount, defines our stringent procedure to set upper limits on $\langle\sigma v\rangle$. In practice, a $\chi^2$ test is performed by comparing the data set $\{M_i\}$, with errors $\{\Sigma_i\}$, to the best fit background (with the normalizations as free parameters) including a DM induced spectrum for a given $\langle\sigma v\rangle  \geq 0$. 
%In practice, a $\chi^2$ test is performed by comparing the data $\{M_i\}$, with errors $\{\Sigma_i\}$, to the best fit background (with the normalizations as free parameters) with a superimposed DM induced spectrum with a fixed $\langle\sigma v\rangle$.  %($\langle\sigma v\rangle \geq 0$)
Our upper limits are then obtained when increasing $\langle\sigma v\rangle$ forces $\chi^2$ to deviate from its best fit value  by more than $\Delta\chi^2 = n^2$. Again the 90, 95 and 99.999\% upper confidence regions are when $n =1.28,1.64$ and 4.3, respectively.

%%%%%%%%%%%%%%%%%%%%%%%%%%%%%%%%%%%%%%%%%%%%
\section{Results and Discussions} \label{results}
\begin{figure}[t]
\centerline{\includegraphics[width=0.99\columnwidth]{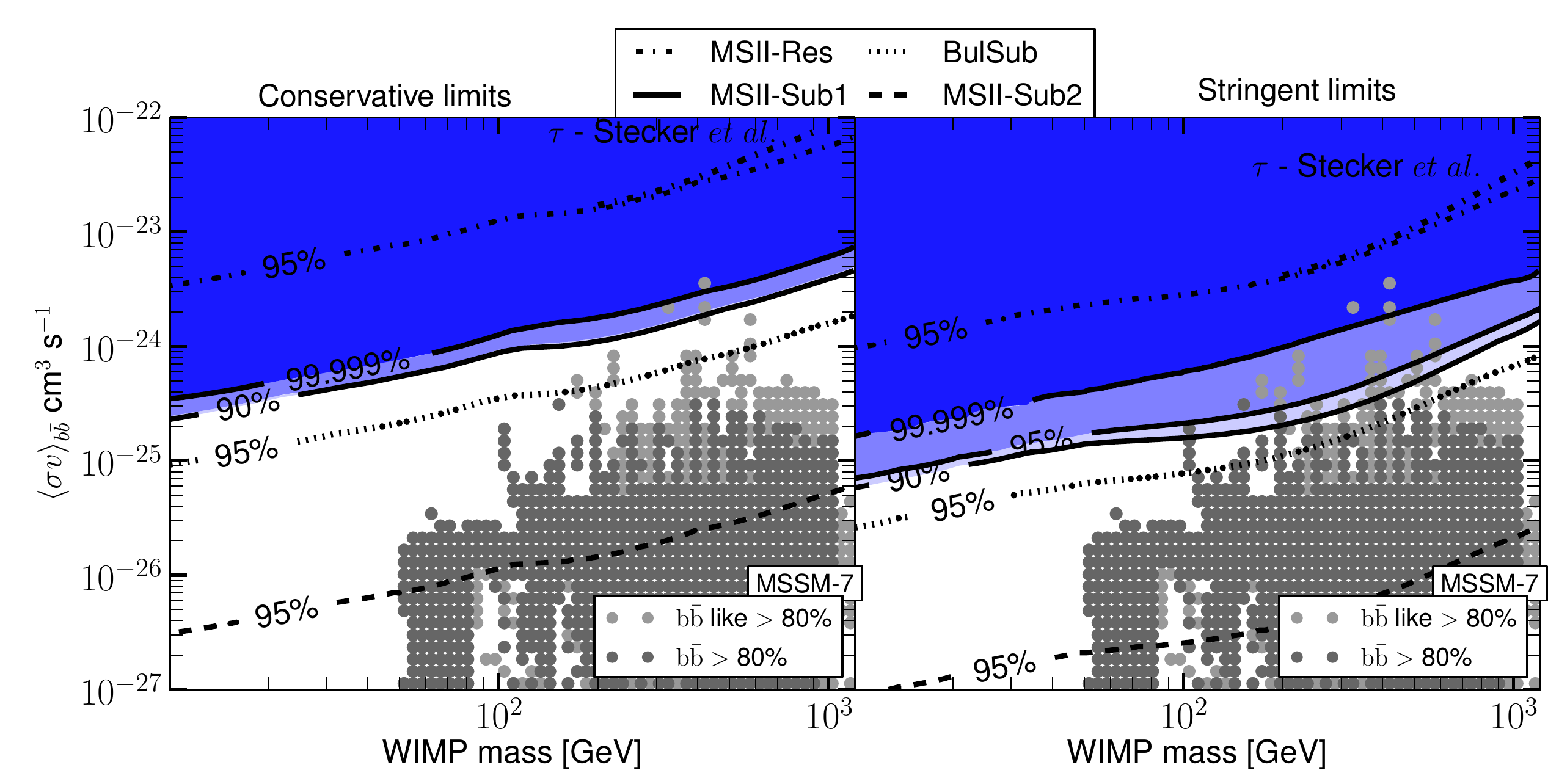}}
 \caption[]{Cross section  $\langle\sigma v\rangle$ limits  on dark matter annihilation into $b\bar{{b}}$ final states. The blue regions mark the (90,~95,~99.999)$\%$ exclusion regions in the MSII-Sub1 $\Delta ^2 (z)$ DM structure scenario  (and for the other structure scenarios only $95\%$ upper limit lines). The absorption model in Gilmore {\it et al.}\ \cite{Gilmore:2009zb} is used, and the relative effect if instead using the Stecker {\it et al.}\ \cite{Stecker:2008fp} model is illustrated by the upper branching of the dash-dotted line in the MSII-Res case. Our {\it conservative} limits are shown on the left and the {\it stringent} limits on the right panel. The grey regions show a portions of the MSSM7 parameter space where the annihilation branching ratio into final states of $b\bar{{b}}$ (or $b\bar{{b}}$ like states) is $> 80\%$. See main text for more details.} \label{ExBB_fig}
\end{figure}

\begin{figure}[t]
\centerline{ \includegraphics[width=0.99\columnwidth]{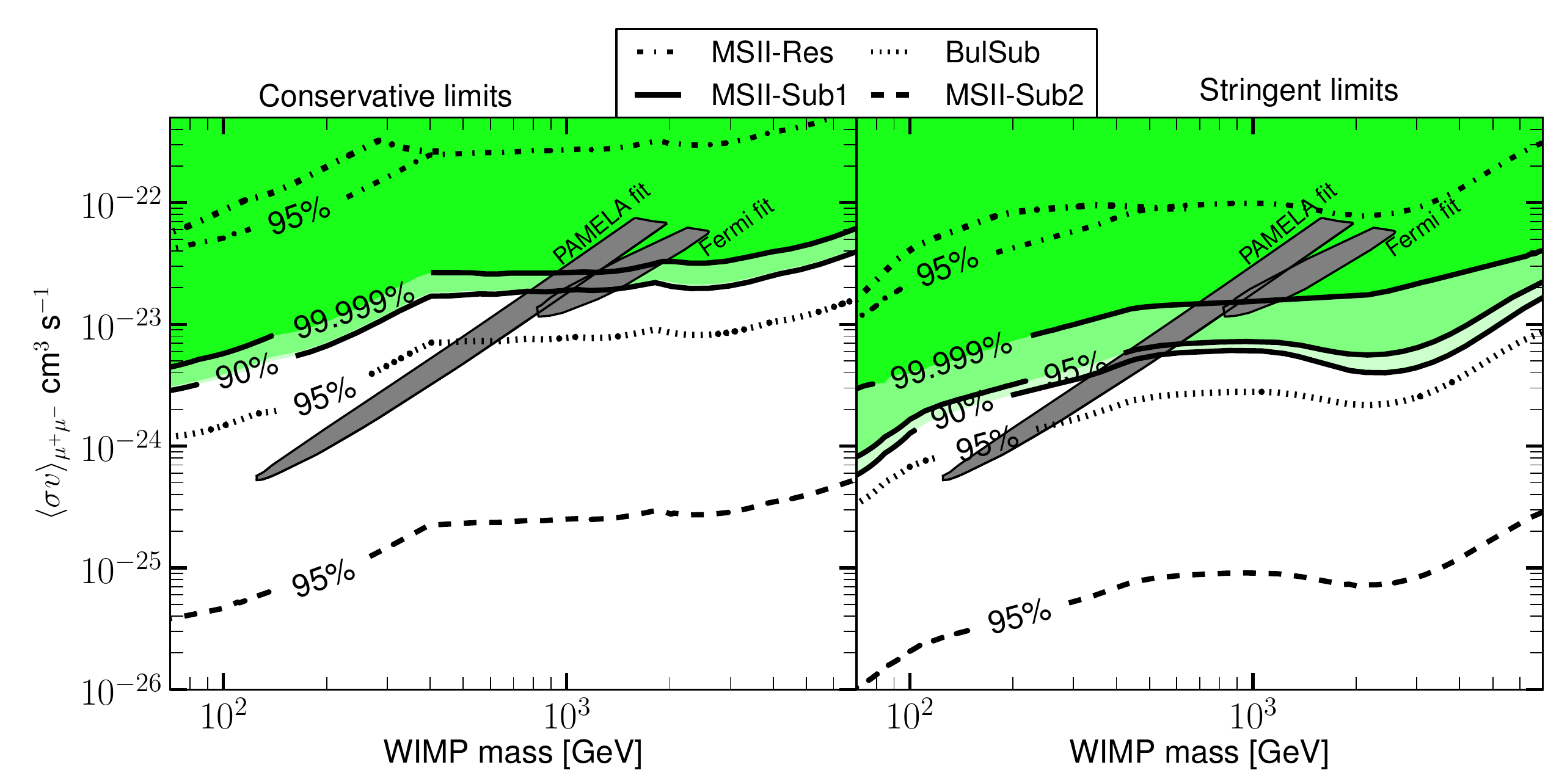} }
 \caption[]{Cross section  $\langle\sigma v\rangle$ limits  on dark matter annihilation into $\mu ^+ \mu^-$ final states. The green regions mark the (90,~95,~99.999)$\%$ exclusion regions in the MSII-Sub1 $\Delta ^2 (z)$ DM structure scenario  (and for the other structure scenarios only $95\%$ upper limit lines). The layout of the figure is otherwise the same as in figure \ref{ExBB_fig}. Note that the Stecker {\it et al.}\ \cite{Stecker:2008fp} absorption model affects the lower DM mass limits since they are set by the high energy FSR part of the DM spectrum. The two gray contours show the best fit regions for a WIMP explanation to the local electron and positron spectra measured by Fermi-LAT and PAMELA.} \label{ExIC_fig}
\end{figure}

\begin{figure}[t]
\centerline{\includegraphics[width=0.99\columnwidth]{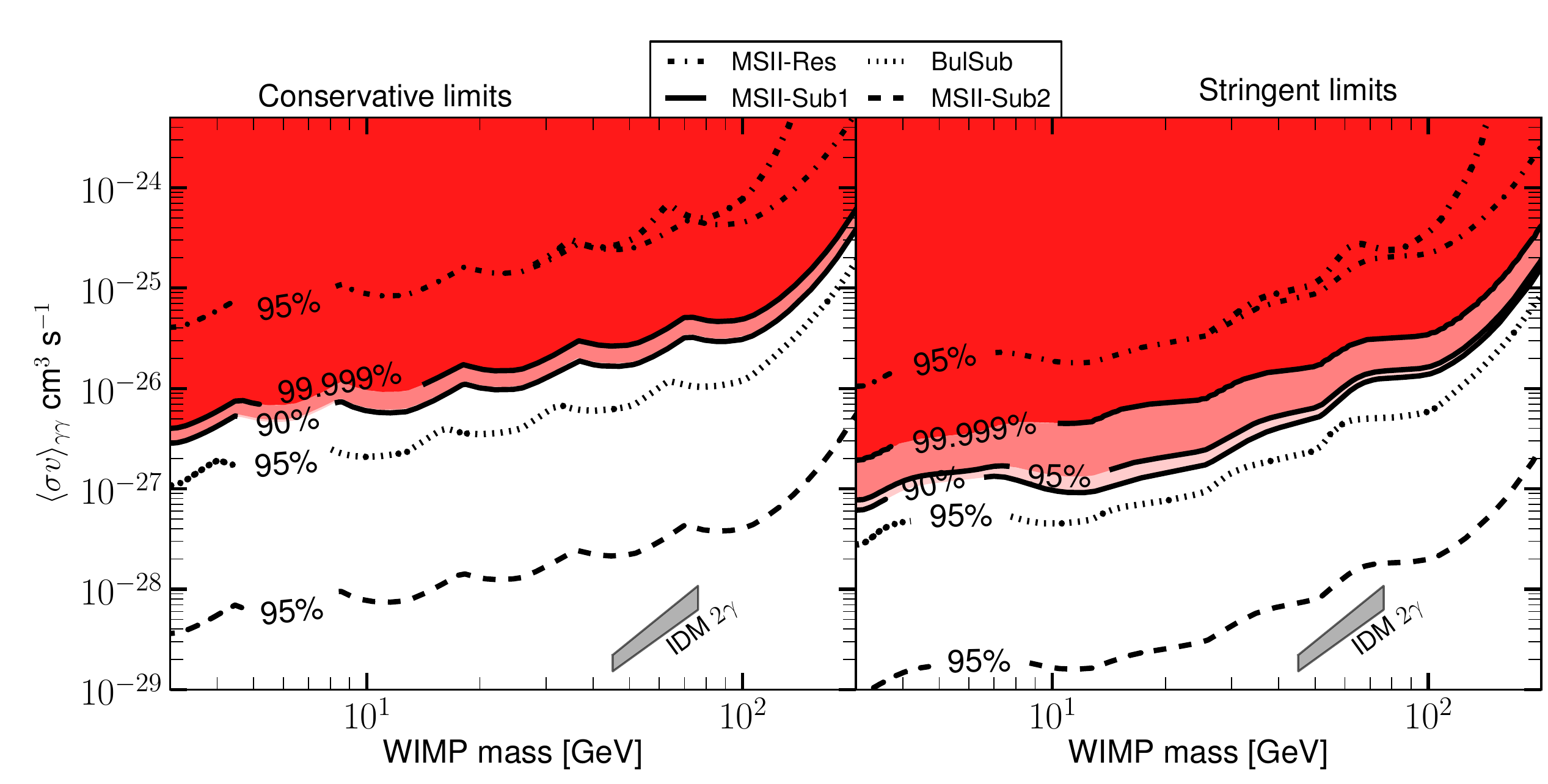}}
\caption[]{Cross section  $\langle\sigma v\rangle$ limits  on dark matter annihilation into two photons. The red regions mark the (90,~95,~99.999)$\%$ exclusion regions in the MSII-Sub1 $\Delta ^2 (z)$ DM structure scenario (and for the other structure scenarios only $95\%$ upper limit lines). The layout of the figure is the same as in figure \ref{ExBB_fig}. Also shown is the IDM parameter space from \protect\cite{Gustafsson:2007pc}.} 
%The wiggles in the limits are caused by the specific energy binning of the IGRB spectrum.
\label{ExLine_fig}
\end{figure}

Figures  \ref{ExBB_fig}, \ref{ExIC_fig} and \ref{ExLine_fig} show the upper cross section limits for our, in section \ref{models}, considered annihilation channels. 
%In figures  \ref{ExBB_fig}, \ref{ExIC_fig} and \ref{ExLine_fig} we present upper cross section limits for our, in section \ref{models}, considered annihilation channels. 
Upper 95\% confidence limits are shown in each of the four DM structure evolution scenarios (MSII-Res, MSII-Sub1, MSII-Sub2 and BulSub), and with additional (90,99.999)\% confidence levels indicated in the MSII-Sub1 case. The outcome from the {\it conservative} and the {\it stringent}  limit analysis procedures, described in section \ref{limits}, are shown in the left and right panels of each figure, respectively. In the case of the conservative limits, the 90 and 95 \% confidence level lines turns out to be close to overlapping, and only the 90\% level is then shown.

The limits are derived with the absorption model by Gilmore {\it et al.}\ \cite{Gilmore:2009zb}, as presented in section \ref{sec:EG}. If instead the Stecker {\it et al.}\  absorption model \cite{Stecker:2008fp} is used, the cross section limits get somewhat weaker for some DM masses. The difference between the absorption models becomes only important when the limits are set by the high energy end of the DM spectra ({\it i.e.}\ energies $\gtrsim 50$ GeV). For DM annihilating into heavy quarks or directly into gammas, this happens for high DM masses ($\geq 600 $ GeV), whereas for the `leptonic' models the change in absorption model affects the lower DM mass limits since they are mostly limited by the high energy FSR part of the DM-induced gamma-ray spectra. We quantify the relative impact, from the two absorption descriptions, on our limits by showing two branches of the dash-dotted cross section lines. To not clutter the plots, the effect is only shown for MSII-Res case, but the relative effect is similar for all our $\Delta ^2 (z)$ scenarios.


\medskip
% [Model~1: ~$b \bar{b}$,~ Figure \ref{ExBB_fig}], 
Figure \ref{ExBB_fig} shows the exclusion regions for DM models annihilating to $b{\bar b}$ final states  %[Model~1: $b \bar{b}$] 
together with a parameter scan result for DM candidates in MSSM7\footnote{This is a seven free parameter MSSM model as defined and used in the DarkSUSY package \cite{Gondolo:2004sc}.} space where the branching ratio into final states of $b\bar{{b}}$ is $> 80\%$. We also show `$b\bar{{b}}$ like' states, defined as when more than 80\% are hadronic fragmentation-induced gamma-rays, as is the case for, {\it e.g.}, gauge bosons and quark as the prompt final states. All MSSM7 models are consistent with accelerator constraints and have neutralino thermal relic abundance corresponding to the inferred cosmological DM density by WMAP \cite{WMAP}. 
%The {\it conservative} limits are shown on the left panel and the {\it stringent} limits on the right one. In the case of the conservative limits, the 90 and $95 \%$ confidence level lines overlap (the same lines overlap in figure  \ref{ExIC_fig} and  \ref{ExLine_fig}).

It is not always direct to compare different works on DM annihilation cross section limits; different physics assumptions, different analysis methods and different data sets  are often used. We will anyway make a comparison to a few other DM constraints, as to put our cosmological DM results into context. With the MSII-Sub2 case our cross section limits are among the strongest indirect detection limits presented to date, but this setup is admittedly a WIMP structure scenario that might be overly optimistic.  The structure and substructure description applied in our BulSub scenario as well as the strict analysis procedure is similar to what was used in the Fermi analysis of Galaxy clusters \cite{fermiclusters} and (with the exception of  no additional inclusion of substructure) the Fermi analysis of dwarf  galaxies  \cite{fermidwarfs}, see also \cite{Scott:2009jn}). It is therefore worthwhile to compare those  
analyses with our BulSub scenario with the strict upper limit calculation procedure.
Our $b \bar b$ cross section limits are, in this perspective,  comparable to the ones presented in the Fermi analysis of dwarf galaxies \cite{fermidwarfs} and somewhat stronger than the constraints from galaxy clusters in \cite{fermiclusters}.  For hadronic annihilation channels,  cosmic-rays, especially antiproton data, can provide comparable limits \cite{Donato:2008jk}. Such limits are, however,  associated with additional uncertainties due the uncertainties related to charged particle propagation in the Galaxy.  In the preparation of this paper, Fermi-LAT data was used in \cite{Cirelli:2009dv,Papucci:2009gd} to set cross section limits on Galactic DM induced gamma-rays. In these two papers, their data analysis method is more similar to our conservative analysis approach, and the presented limits are comparable to our conservative MSII-sub1 limits when their Galactic DM halos are described by a smooth Einasto or NFW DM density profile.  As mentioned, most hadronic channels are very similar in their gamma-ray production. To within a factor of two  (if final states are not very close to, or below, production thresholds) our cross section limits are also valid for prompt annihilation into the  heavy gauge bosons, the other standard model quarks, gluons, as well as into the leptonic $\tau^+\tau^-$ channel. 
%Compared to the one year estimates in the pre-launch paper the actual (stringent) limits set in the first year of Fermi are worse, mostly due to the overly optimistic assumptions made on the size of the error bars in the pre-launch paper.

\medskip
%[Model~2: ~$\mu^+ \mu^-$,~ 
Figure \ref{ExIC_fig} shows the exclusion region for the leptonic DM model, together with the best fit region for this model to the PAMELA and Fermi-LAT positron and electron data. The sharp upper endings of the gray best fit regions come from the constrain to not overshoot HESS data \cite{Collaboration:2008aaa}. Both the best fit regions and the exclusion regions for all our discussed DM scenarios are calculated in a self-consistent way, modulo minor corrections. Below a DM mass of about 500 GeV, the limits on these models are determined by the FSR signal at the high-energy end of the DM spectra, see figure \ref{spec_fig}, and therefore depend more substantially on the choice of the absorption model. We note here that this conclusion holds even if one considers the constraints that the low energy COMPTEL \cite{Weidenspointner:2000aq} and EGRET \cite{Sreekumar:1997un,Strong:2004ry} data would pose on the first (IC) peak in the spectra. The difference between the Stecker {\it et al.}\ \cite{Stecker:2008fp} and the Gilmore {\it et al.}\ \cite{Gilmore:2009zb} absorption model results in a difference in the FSR signal calculated in the two cases by a factor $\lesssim 2$, and affects our limits correspondingly. %We also note the IC peak for DM masses $\leq 800$ GeV is at energies below the Fermi range, and could potentially be better constrained by lower energy COMPTEL/EGRET data. However, we find that the the larger errorbars of COMPTEL make limits less constraining for masses $\leq 300$ GeV, while EGRET sets the same limits as found by Fermi, in the mass range between 300 GeV and 800 GeV. 

As discussed, prior to this work many leptonic DM models adjusted to fit the PAMELA and Fermi data were already in tension with a wide range of experimental studies of  gamma and radio signals \cite{Bergstrom:2008ag,Bertone:2008xr,Regis:2008ij,Cirelli:2009dv,Papucci:2009gd,Crocker:2010gy}, as well as neutrinos \cite{Hisano:2008ah}, from the inner Galaxy, Big Bang Nucleosynthesis \cite{Hisano:2009rc,Jedamzik:2009uy}
%, the integrated cosmological flux of IC photons compared to EGRET data \cite{Profumo:2009uf,Belikov:2009cx}
, and the non-observation of distortions of the cosmic microwave background \cite{Padmanabhan:2005es,Galli:2009zc,Slatyer:2009yq,Zavala:2009mi, Cirelli:2009bb}.  In this context, we want to point out that our limits have a weak dependence on the DM density in the very inner part of the Milky Way (likewise the limits based on Big Bang Nucleosynthesis and non-distortions of the cosmic microwave background).
Also our moderate MSII-Sub1 (at least in the stringent analysis case) and the BulSub DM 
scenarios exclude models that are most favored by the PAMELA/Fermi measurement. The stringent BulSub limit is somewhat stronger than the most optimistic limits set by the Fermi-LAT observation of dwarfs spheroidal galaxies \cite{fermidwarfs} and galaxy clusters \cite{fermiclusters}.  Similar to the cluster analysis, and in contrast to the constraints placed by the non-detection of dwarf spheroidal galaxies, our limits do not rely on the modeling of cosmic ray electron and positron transport and diffusion in DM halos. In the two recent papers \cite{Cirelli:2009dv,Papucci:2009gd}, mentioned previously in this section, they find cross section limits on the $\mu^+\mu^-$ channel from Fermi-LAT data that are fairly similar to our MSII-sub1 stringent limit. Although their limits do not require strong signal enhancements in the inner Galaxy, they still also have uncertainties related to the diffusion modeling \cite{Cirelli:2009dv,Papucci:2009gd}.

For the case of a $e^+e^-$ annihilation channel the constraints would become stronger by a factor of a few compared to $\mu^+\mu^-$, unless it is the FSR that is the constrained process, then the limits are only stronger by a factor $\sim \ln{4 m_{DM}^2 m_e^{-2}} / \ln{4 m_{DM}^2 m_\mu^{-2}}  \lesssim 2$.  For multi lepton final states, such as $\mu^+\mu^-\mu^+\mu^-$, the limits get instead weaker.  

 %Prior to our work, the leptonic DM models were already in tension with wide range of experiments, see for example cite{Bertone:2008xr,Cirelli:2009bb,Galli:2009zc,Bergstrom:2008ag}. Our moderate constraints (MSII-Sub1) are comparable with the most optimistic limits set from the Fermi-LAT observation of Dwarfs Galaxies cite{fermidwarfs}, with our MSII-Sub2 limits ruling out an interesting parameter region by a large margin. However, the uncertainties involved in this analysis prevent us from claiming the full exclusion of the interesting parameter region.

\medskip
%[Model~3: ~$\gamma \gamma$,~ 
Figure \ref{ExLine_fig} shows the cross section limits for DM models annihilating to $\gamma \gamma $ states, together with the gamma-line strength expected in the inert doublet model \cite{Gustafsson:2007pc}. The cross section limit on a DM particle, with mass of $m_{\mathrm DM}$, annihilating instead into a photon and a heavy X particle, with mass $m_{\mathrm X}$, is given by
\begin{equation}
\langle \sigma v \rangle^{ m_{\rm DM}}_{\gamma \rm X, {\rm limit}} = 
2 \frac{m_{\rm DM}^2}{{m'}_{\rm DM}^2} \times \langle \sigma v \rangle^{m'_{\rm{DM} }}_{\gamma \gamma, {\rm limit}}\,, \;\;\hbox{where $m'_{\rm DM} = \frac{m_{\rm DM}}{2} \left(1+\sqrt{1+\frac{m_{\rm X}^2}{m_{\rm DM}^2}} \right) $}.
\end{equation}
The annihilation limit $\langle \sigma v \rangle^{ m_{\rm{DM}} }_{\gamma \gamma, {\rm limit}}$ is directly read from figure \ref{ExLine_fig} at the corresponding WIMP mass $m'_{\rm DM}$. For this expression to be exactly valid, the $X$ particle is assumed to be stable, as otherwise that produce an intrinsic broadening of monochromatic lines. 

Our monochromatic line constraints allows us only to constrain scenarios that predict very strong gamma-lines, such as those presented in \cite{Jackson:2009kg}.  Whereas models with relative strong gamma lines as, {\it e.g.}, in the inert doublet model \cite{Gustafsson:2007pc} or in \cite{Bertone:2009cb} are  still below the sensitivity of the Fermi-LAT isotropic measurement. Even if our most optimistic MSII-Sub2 stringent analysis still present much stronger annihilation cross section constraints than the dedicated line search of the Fermi-LAT collaboration \cite{fermilines}, the more solid upper limit from the MSII-Sub1DM structure scenario are significantly weaker. 
Comparing to the earlier works \cite{Bergstrom:2001jj,Ullio:2002pj} on cosmological DM induced gamma-lines, we find a missing factor (1 GeV/$E_0$) in their presented gamma-line signals. This is accordingly reflected in the substantially weaker flux limits we obtain than anticipated from these papers.

%We also note that our line fluxes differ by a factor (1 GeV/$E_0$) compared to (we believe erroneous) calculation results presented in earlier works \cite{Bergstrom:2001jj,Ullio:2002pj}. This disagreement in the technical calculation is also reflected in the substantially weaker flux %limits we obtain than anticipated from these papers.

We note that the used a binned Fermi-LAT spectrum introduces a wiggly behavior in the upper limits for the case of DM annihilation into monochromatic gamma-rays. This is because the narrow signal is sometimes split between two energy bins.  The sharp rise in the upper limits above 100 GeV is due to the fact that the Fermi-LAT measured spectrum ends there, and that DM induced gamma-lines at high redshifts do not produce very strong signal at lower observed energies because of strong absorption at these energies.

%%%%%%%%%%%%%%%%%%%%%%%%%%%%%%%%%%%%%%%%%%%%
\section{Summary and Outlock} \label{summary}
By using the isotropic gamma-ray background measurement of the Fermi-LAT \cite{egbpaper},  we have set upper limits on the cross section (at 90, 95 and 99.999\% confidence level) for cosmological DM annihilating into different final states that induce high-energy gamma-rays. Such cross section limits have associated uncertainties, with the largest uncertainty stemming from the modeling of the evolution of DM structure and  substructure. The difficulty of estimating the isotropic background to the cosmological DM annihilation signal further increases the uncertainty in limits. Given these uncertainties, our most conservative and most optimistic limits on cross sections span three orders of magnitude. While the most conservative constraints barely reach exclusion of theoretically discussed DM cross sections, more optimistic descriptions of the DM halos and subhalos would allow to exclude all (or an interesting fraction of) such models. This demonstrates the potential the isotropic signal has to observe a DM signal, but also the importance to better understand the DM structure formation in order to reduce uncertainties and enable firm constraints on electroweak properties of DM particles. Better knowledge of the conventional extra galactic astrophysical contributions and of the galactic gamma-ray foreground model would also further improve the understanding and interpretation of the IGRB.
If the main part of the Fermi-LAT measured IGRB originates from unresolved extragalactic objects, then, by continuing to resolve astrophysical contributions, the IGRB flux could be reduced. This effect would then be to lower the measured IGRB flux during the Fermi mission, which, together with improved statistics and extended energy range, would allow to probe a larger fraction of the DM parameter space. 

The DM signal from our own Galaxy (both the more isotropic signal from Galactic DM substructures, and the anisotropic signal expected from the smooth Galactic main halo) could also make an important contribution. It is therefore expected that a simultaneous analysis of morphology and gamma-ray energy spectra, including conventional and DM Galactic as well as extragalactic signals, could be beneficial in the search of DM signals. 

This paper also investigated the importance of the transparency of the Universe to high energy gamma-rays on the perceived DM signals. We find that, in the cases when the limits are set by the high energy part of the DM annihilation spectra (for energies $\gtrsim 50$ GeV), there is notable difference between existing models of the optical depth.

%%%%%%%%%%%%%%%%%%%%%%%%%%%%%%%%%%%%%%%%%%%%
\section*{Acknowledgements}
The \textit{Fermi} LAT Collaboration acknowledges generous ongoing support from a number of agencies and institutes that have supported both the development and the operation of the LAT as well as scientific data analysis. These include the National Aeronautics and Space Administration and the Department of Energy in the United States, the Commissariat \`a l'Energie Atomique and the Centre National de la Recherche Scientifique / Institut National de Physique Nucl\'eaire et de Physique des Particules in France, the Agenzia Spaziale Italiana and the Istituto Nazionale di Fisica Nucleare in Italy, the Ministry of Education, Culture, Sports, Science and Technology (MEXT), High Energy Accelerator Research Organization (KEK) and Japan Aerospace Exploration Agency (JAXA) in Japan, and the K.~A.~Wallenberg Foundation, the Swedish Research Council and the Swedish National Space Board in Sweden.
\\ \indent
Additional support for science analysis during the operations phase is gratefully acknowledged from the Istituto Nazionale di Astrofisica in Italy and the and the Centre National d'\'Etudes Spatiales in France.
\\ \indent
We would like to thank Lars Bergstr\"om, Joakim Edsj\"o, Jennifer Siegal-Gaskins, Miguel Pato, Anders Pinzke, Rudy Gilmore and Jesus Zavala for useful discussions. J.~Conrad is a Royal Swedish Academy of Sciences Research Fellow supported by a grant of the Knut and Alice Wallenberg foundation. M.~Gustafsson and G.~Zaharijas also acknowledge support from the National Institute of Nuclear Physics and the European Union FP6 Marie Curie Research \& Training Network `UniverseNet' (MRTN-CT-2006-035863).

%%%%%%%%%%%%%%%%%%%%%%%%%%%%%%%%%%%%%%%%%%%%
%%%%   Appendix
%%%%%%%%%%%%%%%%%%%%%%%%%%%%%%%%%%%%%%%%%%%%
%\appendix

%%%%%%%%%%%%%%%%%%%%%%%%%%%%%%%%%%%%%%%%%%%%
%%%%   References
%%%%%%%%%%%%%%%%%%%%%%%%%%%%%%%%%%%%%%%%%%%%

%%%%%%%%%%%%%%%%%%%%%%%%%%%%%%%%%%%%%%%%%%%%
\end{document}